\documentclass[aps,prb,twocolumn,a4paper,showpacs,floatfix,superscriptaddress]{revtex4}

\usepackage{amsmath, amssymb,mathrsfs}
\usepackage{graphicx}
\usepackage{hyperref}
\usepackage{braket}
\usepackage{nicefrac}
\usepackage[squaren]{SIunits}
\usepackage[]{subfig}

\newcommand{\e}{\mathrm{e}}
\newcommand{\ii}{\mathrm{i}}
\newcommand{\Lh}{\nicefrac{L}{2}}

\newcommand{\upd}{\mathrm{d}}

\renewcommand{\div}{\mathrm{div}\,}

\newcommand{\vek}[1]{\mathbf{#1}}
\newcommand{\vekh}[1]{\hat{\mathbf{#1}}}

\newcommand{\vq}{\vek{q}}
\newcommand{\qpara}{\vek{q}_{\parallel}}
\newcommand{\qperp}{q_{\perp}}

\newcommand{\gk}{^{\gtrless}}
\newcommand{\kg}{^{\lessgtr}}

\newcommand{\med}{_{\rm med}}
\newcommand{\vac}{_{\rm vac}}

\newcommand{\ext}{_{\mathrm{ext}}}
\newcommand{\ind}{_{\mathrm{ind}}}

\newcommand{\Dret}{D^{\mathrm{ret}}}
\newcommand{\Doret}{D_0^{\mathrm{ret}}}

\newcommand{\Dadv}{D^{\mathrm{adv}}}

\newcommand{\Doadv}{D_0^{\mathrm{adv}}}

\newcommand{\Dm}{D\med}
\newcommand{\Dv}{D\vac}

\renewcommand{\Re}{\mathrm{Re}\,}
\renewcommand{\Im}{\mathrm{Im}\,}

\newcommand{\jext}{\vek{j}_{\mathrm{ext}}}

\newcommand{\epstret}{\varepsilon_T^{\rm ret}}
\newcommand{\epstreti}{\varepsilon_T^{\rm ret,-1}}

\newcommand{\epstadvi}{\varepsilon_T^{\rm adv,-1}}

\newcommand{\ia}{_{\lambda \vq}}
\newcommand{\ib}{_{\lambda' \vq'}}

\newcommand{\iab}{_{\lambda \vq \lambda' \vq'}}

\newcommand{\iai}{_{\lambda \vq,i}}
\newcommand{\ibj}{_{\lambda' \vq',j}}
\newcommand{\iaj}{_{\lambda \vq,j}}

\newcommand{\iq}{_{\vq}}

\newcommand{\cc}{\mathrm{\ldots\,c.c.\,\ldots}}

\begin{document}

\title{Green function approach to scattering of nonclassical light by bounded media}
\author{F.~Richter}
\email{felix.richter2@uni-rostock.de}
\affiliation{Institut f\"ur Physik, Universit\"at Rostock, 18051 Rostock, Germany}
\author{D.~Yu.~Vasylyev}
\email{dmytro.vasylyev@uni-rostock.de}
\affiliation{Institut f\"ur Physik, Universit\"at Rostock, 18051 Rostock, Germany}
\affiliation{Bogolyubov Institute for Theoretical Physics, NAS of Ukraine, UA-03680 Kiev, Ukraine}
\author{K.~Henneberger}
\affiliation{Institut f\"ur Physik, Universit\"at Rostock, 18051 Rostock, Germany}

\begin{abstract}
We show that using the properties of the photon Green's function one can successfully describe the propagation of arbitrary nonclassical optical radiation through structured materials. In contrast to the similar input-output approach, our method is not restricted neither to the spatial homogeneous matter nor to the specific direction of light beam incidence or outcoupling. Several quantum states of light are addressed in detail, such as Fock, Glauber, and different squeezed light states, and their Green functions are given.
Furthermore, it is demonstrated how scattering of light at a slab can be described with 
Poynting's energy flow vector, and how experimental setups can be modeled.
\end{abstract}


\maketitle 

\section{Introduction}\label{introduction}
The nonequilibrium photon Green function (GF)
\begin{multline}
\label{eq-def-pgf-gk}
\ii \hbar \mu_0 D_{ik}^>(1,2) = \ii \hbar \mu_0 D_{ki}^<(2,1) \\ 
=  \braket{\hat A_i(1)\hat A_k(2)} - \braket{\hat A_{i}(1)}\braket{\hat A_{k}(2)} .
\end{multline}
is defined in terms of expectation values of correlations of the vector potential operator $\vekh A$. Its average $\vek A = \Braket{\vekh A}$ obeys Maxwell's potential equation (wave equation)
\begin{align}
\label{eq-maxwell-pot-keldysh}
\square \vek{A}(\vek r, t) = -\mu_0 \left[ \vek j\ind(\vek r, t) + \jext(\vek r, t)\right]\,.
\end{align}

The photon GF was shown in Ref.\ \onlinecite{Henneberger2009} to exhibit an exact splitting into two components
\begin{subequations}
\begin{align}
\label{eq-Dgk-expl}
D\gk &= \Dm\gk + \Dv\gk \, ,\\
\label{eq-Dmed}
\Dm\gk &=  \Dret  P\gk \Dadv \, ,\\
\label{eq-Dvac}
\Dv\gk &=  \epstreti  D\gk_0 \, \epstadvi \, ,
\end{align}
\end{subequations}
which can clearly be identified as contributions arising from electronic generation and recombination processes in the medium ($P\gk$) in the case of the \textit{medium-induced contribution} $D\med\gk$ or, respectively, from the electromagnetic state in the free space (vacuum) surrounding the medium, represented by the \textit{vacuum Green function} $D\gk_0$, in the case of the \textit{vacuum-induced contribution} $D\vac\gk$.

Here, $\epstret$ is the transverse dielectric tensor, which describes renormalization of a freely evolving vector potential
\begin{align}
\label{eq-def-aext}
\square \vek{A\ext}(\vek r, t) &= -\mu_0 \jext(\vek r, t) \,
\end{align}
due to the presence of the medium to the actual vector potential $\vek A$ in linear approximation according to
\begin{align}
\label{eq-epstreti-a}
\vek A = \epstreti \vek A\ext \,.
\end{align}

If the vacuum GF $D_0\gk$ is given in terms of free plane waves, $D\vac\gk$ can be obtained trivially by replacement of the free waves by renormalized waves $\vek A$.\cite{Henneberger2009}

The splitting of $D\gk$ [Eq.\ \eqref{eq-Dgk-expl}] translates directly to the energy flux in the system, which is given by the Poynting vector, and we can clearly identify the energy flux contribution caused by the state of light in the free space, i.e., its vacuum-induced contribution.\cite{Henneberger2009}

In Sec.\ \ref{sec-preparation}, we will derive and discuss the vacuum Green function and the respective vacuum-induced contribution to the full Green function for arbitrary states, following closely the approach sketched in Ref.\ \onlinecite{Henneberger2009}. In particular, Fock, Glauber, and different squeezed light states are addressed.

In Sec.\ \ref{sec-energyflow}, the vacuum-induced energy flow in the slab geometry will be developed in more detail to allow eventually for the description of scattering in optical setups by Green functions. Modelling of beams splitters and photon detection is briefly addressed.

The advantage of this approach is the exact consideration of the spatial inhomogeneities inherent to bounded media systems. Since the Green functions are globally defined, piecewise definition of photon operators is not necessary.

\section{Vacuum Green function for arbitrary quantum states}
\label{sec-preparation}
The vacuum GF $D_0\gk$ consists of the inevitable spontaneous ground state fluctuations of the electromagnetic vacuum, which we will denote by $D\gk_{\rm 0,sp}$, and a stimulation $D\gk_{\rm 0,stim}$ which is given by the preparation of the free space to a certain electromagnetic state.

The complete fluctuations in the free space can be written as
\begin{equation}
\label{eq-Dgk0-def}
D\gk_0(1,2) =  D\gk_{\rm 0,sp}(1 - 2) + D\gk_{\rm 0,stim}(1,2). 
\end{equation}
Note that always $D^>_{\rm 0,stim}(1,2)=D^<_{\rm 0,stim}(1,2)$, since the vacuum spectral function 
\begin{align}
  \hat D_0 = D_0^> - D_0^< = D_{\rm 0,sp}^> - D_{\rm 0,sp}^<  = \Doret - \Doadv
\end{align}
is a unique and universal function,\cite{Henneberger2009} such that the $\gk$ indication may be omitted in the following.

\subsection{Normal-mode expansion in a mixed state}
\label{subsNormMode}
We apply in $D\gk_0(1,2)$ a normal mode expansion for the freely evolving vector potential operator\cite{Vogel2006} 
\begin{multline}
\vekh{A}_0(\vek r) =  \\\sum_{\lambda \vek q} \sqrt{\frac{\hbar}{2\varepsilon_0 c q V}} \vek e_{\lambda \vek q} \left( \hat{a}_{\lambda \vek q} \e^{\ii \vek q \vek r} + \hat{a}_{\lambda \vek q}^+   \e^{-\ii \vek q \vek r} \right) \, ,
\end{multline}
where $ \vek e_{\lambda \vek q }$ denote the transverse polarization vectors and $\hat{a}_{\lambda \vek q}$ ($\hat{a}_{\lambda \vek q}^+$) are photon annihilation (creation) operators. Following the definition \eqref{eq-def-pgf-gk}, one obtains the following structure:
\begin{widetext}
  \begin{subequations}
  \label{eq-d0gk-general}
   \begin{eqnarray}
D^<_{0,ij}(1,2) &=& \frac{c}{2\ii V}\sum\ia \sum\ib \frac{1}{\sqrt{qq'}} 
\left\{C_{n,\lambda\vq\lambda'\vq'}^<\,F\iai(1) F\ibj^*(2)
+C_{n,\lambda\vq\lambda'\vq'}^>\,F\iai(1) F\ibj^* (2)\right.\nonumber
\\ 
&&\left.+\left(C_{a,\lambda\vq\lambda'\vq'}\, F\iai(1) F\ibj(2) + \cc\right)\right\} \\
D^>_{0,ij}(1,2) &=& \frac{c}{2\ii V}\sum\ia \sum\ib \frac{1}{\sqrt{qq'}} 
\left\{C_{n,\lambda'\vq'\lambda\vq}^>\,F\iai(1) F\ibj^*(2)
+C_{n,\lambda'\vq'\lambda\vq}^<\,F\iai(1) F\ibj^* (2)\right.\nonumber
\\
&&\left.+\left(C_{a,\lambda\vq\lambda'\vq'}\, F\iai(1) F\ibj(2) +\cc\right)\right\} 
   \end{eqnarray}
  \end{subequations}
\end{widetext}
where 
\begin{equation}
\label{eq-free-prop}
\vek F\ia(1) = \vek e\ia \, \exp\left[\ii (\vq \vek r_1 - c|\vq|t_1 )\right]\,
\end{equation} 
describe classical plane waves with polarization $\vek e\ia$ and wave vector $\vq$. Note that the second term is not the complex conjugate of the first to due the index switch.

The terms $\propto C_n\gk$ are \textit{normal} terms, which depend on the difference variables $(1-2)$ only and are thus homogeneous. In contrast, the \textit{anormalous} terms $\propto C_a$ depend on $(1+2)$ and thus pertain to inhomogeneous systems.
The prefactors $C$ are expectation value differences of correlated and uncorrelated creation and annihilation operators in an arbitrary base:
\begin{subequations}
\label{eq-def-expval}
\begin{align}
  C_{n,\lambda\vq,\lambda'\vq'}^> 
      &= \braket{\hat a\ia   \hat a\ib^+} - \braket{\hat a\ia}  \braket{\hat a\ib^+}\\
  C_{n,\lambda\vq,\lambda'\vq'}^< 
      &= \braket{\hat a\ia^+ \hat a\ib  } - \braket{\hat a\ia^+}\braket{\hat a\ib}\\
  C_{a,\lambda\vq,\lambda'\vq'} 
      &= \braket{\hat a\ia   \hat a\ib  } - \braket{\hat a\ia}  \braket{\hat a\ib}
\end{align}
\end{subequations}

Application of the commutation relation $[\hat a\ia, \hat a\ib^+] = \delta_{\iab}$ in the expectation values above (necessary to obtain normal order in $C_n^>$) reveals the ground state contribution to $D_0\gk$,
\begin{multline}
  \label{eq-d0gk-sp}
  D^>_{{\rm 0,sp},ij}(1 - 2) = D^<_{{\rm 0,sp},ij}(2 - 1)\\
= \sum\ia \frac{c}{2 \ii V q} F_{\lambda\vq,i}(1) F_{\lambda\vq,j}^*(2)\,,
\end{multline}
since, in any (mixed) quantum state described by a statistical operator $\hat \rho$,
\begin{multline}
  \label{eq-aa-commut}
  \braket{\hat a\ia   \hat a\ib^+} =  \mathrm{Tr}\left\{\hat\rho\,\hat a\ia \hat a^+\ib \right\}\\ = \delta\iab + \mathrm{Tr}\left\{\hat\rho\,\hat a^+\ib \hat a\ia \right\}\,.
\end{multline}

After separation of $D\gk_{\rm 0,sp}$, the remaining terms in Eq.\ \eqref{eq-d0gk-general} constitute $D_{\rm 0, stim}$. They are mutual complex conjugates, and the latter can be written in a more compact form by definition of the componentless prefactor
\begin{align}
\label{eq-def-expval-stim}
  C_{n,\lambda\vq\lambda'\vq'} &= \braket{\hat a\ib^+ \hat a\ia} - \braket{\hat a\ia}  \braket{\hat a\ib^+}
\end{align}
as
\begin{widetext}
\begin{align}
\label{eq-dstim-general}
D_{{\rm 0,stim},ij}(1,2) = \frac{c}{2\ii V}\sum\ia \sum\ib \frac{1}{\sqrt{qq'}} 
\left\{C_{n,\lambda\vq\lambda'\vq'}\, F\iai(1) F\ibj^*(2) + C_{a,\lambda\vq\lambda'\vq'}\, F\iai(1) F\ibj(2) + \cc \right\}\,.
\end{align}
\end{widetext}

Using Eq.\ \eqref{eq-Dvac}, the vacuum fluctuations appear renormalized due to the presence of a bounded medium according to\cite{Henneberger2009}
\begin{align}
\label{eq-dvgk-general}
\Dv\gk(1,2) = D_0\gk(1,2; \vek F \to \vek A)\,,
\end{align}
i.e., with $\vek F$ replaced by the effective fields
\begin{align}
\label{eq-prop-prop}
  \vek A\ia(1) &= \epstreti(1,2) \,\vek F\ia(2)\,,\\
  \vek A\ia^*(1) &= \epstreti(1,2) \,\vek F\ia^*(2)\,,
\end{align} 
which describe propagation (i.e., reflection, absorption, and transmission) of a classical plane wave in the presence of a bounded medium. They are normal mode expansions of the vector potential and solutions of Eq.\ \eqref{eq-epstreti-a}.

\subsection{Fock state}
So far we did not make any assumptions on the state of the light in the free space. The case of a multi-mode number (or Fock) state $\ket{\{n\}\ia}$ with the photon population $n\ia = n^<\ia = n^>\ia - 1$ was discussed in Ref.\ \onlinecite{Henneberger2009} already. The result from evaluation of the expectation values \eqref{eq-def-expval} is
\begin{multline}
\label{eq-d0gk-fock}
D\gk_{0,ij}(1-2) =  \sum\ia \frac{c}{2 \ii q} \, 
 \left [   n\gk\ia  \, F\iai(1) F\iaj^*(2) \right. \\
  \left. + n\kg\ia  \, F\iai^*(1) F\iaj(2) \right ]\, .
\end{multline}
Uncorrelated expectation values as well as anormalous contributions vanish ($\bar D_0\gk \equiv D_0\gk$). The GF is homogeneous, and its Fourier domain representation is
\begin{multline}
\label{eq-d0gk-fock-fourier}
D\gk_0(\vek q, \omega) = \\ 
\frac{\pi c}{ \ii q }\sum_\lambda \left[n\ia\gk \delta(\omega-cq) + n\ia\kg \delta(\omega+cq)\right]\,.
\end{multline}
The ground state contribution is contained in the $n^>$ terms and the stimulated part is
\begin{multline}
\label{eq-d0gk-fock-stim}
D\gk_{0,{\rm stim},ij}(1{-}2){=}  \sum\ia \frac{c}{\ii q} \, 
 n\ia \Re \left [F\iai(1) F\iaj^*(2) \right ] \, .
\end{multline}

We will now derive and discuss the vacuum GF $D_0\gk$ for other interesting states.

\subsection{Coherent state}
For a coherent (Glauber) state $\ket{\beta}$ with displacement $\beta$ such that\cite{Vogel2006}
\begin{align}
  \hat a\ket{\beta} = \beta\ket{\beta}\,, \quad \bra{\beta}\hat a^+=\bra{\beta}\beta^* \,, \quad \braket{\alpha|\beta}\neq \delta_{\alpha,\beta}\,,
\end{align}
one finds 
\begin{subequations}
\label{eq-c-glauber}
\begin{align}
C_n &= 0 & C_a &= 0
\end{align}
\end{subequations}

\subsection{Squeezed light states}
A squeezed light state $\ket{\xi,\phi}$ evolves through a finite-time interaction, e.g., in a nonlinear crystal, from a given state $\ket{\phi}$. This evolution is described by the unitary operator $\hat S(\xi)$, the \textit{squeezing operator}: $\ket{\xi,\phi} = \hat S(\xi)\ket{\phi}$. It reads in a generalized form for multimode squeezing\cite{Lo1993,Vogel2006}
\begin{multline}
  \label{eq-def-squeezing-operator}
  \hat S(\xi) = \exp\left[\sum\ia\sum\ib\left(\xi\iab^* \hat a\ia \hat a\ib \right.\right.\\
  \left.\left. - \xi\iab \hat a\ia^+ \hat a\ib^+\right)\right] \,,
\end{multline}
where $\xi$ is a matrix describing the coupling strength between different modes.

The expectation values \eqref{eq-def-expval} can be obtained in a straightforward manner using the relations
\begin{subequations}
\label{eq-def-sas-general}
\begin{align}
\hat S^+\hat a\ia \hat S &= \sum\ib\left(   \mu\iab   \hat a\ib - \nu\iab\hat a^+\ib\right) \\
\hat S^+\hat a\ia^+ \hat S &= \sum\ib\left( \mu^*\iab \hat a\ib^+ - \nu^*\iab\hat a\ib\right)\, ,
\end{align}
\end{subequations}
where $\mu$ and $\nu$ are $\cosh$- and $\sinh$-like power series in $\xi$.\cite{Lo1993} Obviously, this will lead to an additional mode sum in Eq.\ \eqref{eq-d0gk-general}, but the general structure remains.

\subsubsection{Single-mode squeezing, squeezed vacuum}

The case of single-mode or diagonal squeezing, i.e., squeezing of a mode with itself, with $\bar\xi\ia/2 = \xi\iab\delta\iab$, allows us to quickly obtain the GF of the squeezed vacuum. The vacuum GF prefactors for a diagonally squeezed Fock state $\ket{\{\bar{\xi},n\}\ia}$ are
\begin{subequations}
\label{eq-expval-diag-squeezed-fock}
\begin{align}
  C_n\gk(\bar\xi) &= (\mu\ia^2 n\gk\ia + |\nu\ia|^2 n\kg\ia)\delta\iab \\
  C_a(\bar\xi) &= -\mu\ia\nu\ia(n^>\ia + n^<\ia)\delta\iab
\end{align}
with the squeezing strength factors $\mu = \cosh|\bar\xi|$ and $\nu = \sinh|\bar\xi|\exp(\ii \arg \bar\xi)$.
\end{subequations}
The double mode sum in Eq.\ \eqref{eq-d0gk-general} then reduces to a single one. We may rewrite $\mu^2 = |\nu|^2 + 1$ and find
\begin{subequations}
\begin{align}
  C_n &= n\ia + |\nu\ia|^2 + 2|\nu\ia|^2 n\ia \\
  C_a &= \mu\ia\nu\ia + 2\mu\ia\nu\ia n\ia
\end{align}
\end{subequations}
and decompose the $D_{\rm 0,stim}$ this way into three contributions: (i) \textit{normal Fock} contribution $D_{\rm 0,nf}\gk$ [$\propto n$, cf.\ Eq.\ \eqref{eq-d0gk-fock}], (ii) \textit{squeezed vacuum} contribution [$\propto \mu,\nu$ alone], which remains even for vanishing mode population,
\begin{multline}
\label{eq-d0gk-squeezed-vacuum}
D_{{\rm 0,sv},ij}(1,2) = \sum\ia \frac{c}{2 \ii V q}
\left\{|\nu\ia|^2 F\iai(1) F\iaj^*(2) \right. \\
\left. - \mu\ia\nu\ia F\iai(1) F\iaj(2) + \text{c.c.} \right\}\,, 
\end{multline}
and (iii) the \textit{squeezed Fock} contribution $D_{\rm 0,sf}$ [$\propto n, \propto \mu,\nu)$].

This results was already given in Ref.\ \onlinecite{Henneberger2009}. Obviously, vanishing of the squeezed contributions $D_{\rm 0,sv}$ and $D_{\rm 0,sf}$ is assured for $\bar\xi \to 0$. Note also that these two parts are always inhomogeneous, since they contain a $C_a$ term.

\subsubsection{Squeezed vacuum in a mixed state}
Since any mixed state may be expanded into Fock states,
\begin{align}
  \mathrm{Tr}\left\{\hat\rho \Phi \right\} = \sum_j \Braket{n_j|\Phi|n_j} \,,
\end{align}
it is easy to see that the squeezed vacuum GF, Eq.\ \eqref{eq-d0gk-squeezed-vacuum}, will appear in any single-mode squeezed (mixed) state. Evaluating this Fock state expansion for general multimode squeezing reveals a squeezed vacuum GF as well, with the same structure (but different definition of $\mu,\nu$). In conclusion, any squeezed light GF exhibits a squeezed vacuum contribution of the form \eqref{eq-d0gk-squeezed-vacuum}.

Note again that $D_{\rm 0,sv}$ is defined here such that it is free from contributions from the ground state, $D_{\rm 0,sp}$.

\subsubsection{Single-mode squeezed coherent light}

For a diagonally squeezed coherent state $\ket{\{\bar\xi,\beta\}\ia}$, correlated and uncorrelated expectation values largely compensate in $D_{\rm 0,stim}$, and only the squeezed vacuum remains.
\begin{subequations}
\begin{align}
  C_{n,\lambda\vq\lambda'\vq'} &= |\nu\ia|^2\delta\iab \\
  C_{a,\lambda\vq\lambda'\vq'} &= -\mu\ia\nu\ia\delta\iab 
\end{align}
\end{subequations}

\subsubsection{Two-mode squeezing}
In a widely used setup for the generation of squeezed or entangled light, the optical parametric oscillator (OPO), photons are emitted from a nonlinear crystal in pairs with correlated moments $\vq_0 \pm \vq$ around the reference momentum $\vq_0$. It may be regarded as an outer parameter of the system and $\vq$ as a variable. Usually, the squeezing strength $\tilde\xi$ is assumed to be symmetrical around $\vq_0$. It vanishes for $\vq=0$ and has a finite bandwidth ($\xi\ia \to 0$ for $q \to \infty$). 

The squeezing operator can be obtained from Eq.\ \eqref{eq-def-squeezing-operator} by diagonalization:\cite{Vogel2006}
\begin{multline}
  \hat S(\tilde\xi) = \exp\left[\sum\ia\left(\tilde\xi\ia^* \hat a_{\lambda,\vq_0 + \vq} \hat a_{\lambda,\vq_0 - \vq} \right.\right.\\
  \left.\left. - \tilde\xi\ia \hat a_{\lambda,\vq_0 + \vq} ^+ \hat a_{\lambda,\vq_0 - \vq} ^+\right)\right]
\end{multline}
For this case, it is convenient to express the coefficients $\mu,\nu$ by functions
\begin{align}
\mu_\lambda(|\vq_0 - \vq|)=\tilde\mu\ia\,, \quad \nu_\lambda(|\vq_0 - \vq|)=\tilde\nu\ia,
\end{align}
such that the relations \eqref{eq-def-sas-general} read now
\begin{subequations}
\begin{align}
\hat S^+\hat a\ia \hat S &=  \tilde\mu\ia \hat a\ia - \tilde\nu\ia\hat a^+_{\lambda,2\vq_0 - \vq}\,, \\
\hat S^+\hat a\ia^+ \hat S &=  \tilde\mu\ia\hat a\ia^+ - \tilde\nu^*\ia\hat a_{\lambda,2\vq_0 - \vq}\,.
\end{align}
\end{subequations}

In a Fock state, the expectation values are [cf.\ Eq.\ \eqref{eq-expval-diag-squeezed-fock}]
\begin{subequations}
\label{eq-expval-tmc-squeezed-fock}
\begin{align}
  C_n\gk(\tilde\xi) &= (\tilde\mu\ia^2 n\gk\ia + |\tilde\nu\ia|^2 n\kg_{\lambda,2\vq_0-\vq})\delta\iab \\
  C_a(\tilde\xi) &= -\tilde\mu\ia\tilde\nu\ia(n^>_{\lambda,2\vq_0-\vq} + n^<_{\lambda,2\vq_0-\vq})\delta\iab \, ,
\end{align}
\end{subequations}
or, in the compact notation for $D_{\rm 0,stim}$ alone:
\begin{subequations}
\label{eq-expval-tmc-squeezed-fock-dstim}
\begin{align}
  \begin{split}  
  C_n(\tilde\xi) &= (|\tilde\nu\ia|^2 + n\ia + |\tilde\nu\ia|^2 n\ia\\
  &\qquad + |\tilde\nu\ia|^2 n_{\lambda,2\vq_0-\vq})\delta\iab
  \end{split}\\
  C_a(\tilde\xi) &= \left(-\tilde\mu\ia\tilde\nu\ia - 2\tilde\mu\ia\tilde\nu\ia n_{\lambda,2\vq_0-\vq}\right)\delta\iab \, .
\end{align}
\end{subequations}
Again, the GF decomposes into three components. Comparing them to the diagonally squeezed Fock state, one finds that, except for $\mu \to \tilde\mu$, only the $D_{\rm 0,sf}$ component differs.

\subsection{Propagation of squeezed light}

The propagated squeezed light correlations (and their propagated components) may be obtained as before according to Eq.\ \eqref{eq-dvgk-general}. Their complex prefactors $C_n,C_a$ and inhomogeneities introduced by squeezing remain.

\section{Energy flow with nonclassical light}
\label{sec-energyflow}

We will now consider energy flow and light scattering with the help of Poynting's theorem,\cite{Richter2008} $\partial U/\partial t + \div \vek S = -\vek j \vek E$, following closely the lines of Ref.\ \onlinecite{Henneberger2009} but extending and refining the approach presented there.

Up to now, the results presented in this work are valid for arbitrarily inhomogeneous and instationary systems. For an easier discussion and better comparison with former work, 
we will now regard a system in \textit{slab geometry}, i.e., TE-polarized light propagating along the $x$ axis through a linear medium slab of thickness $L$ which is infinitely extended in the $y$-$z$ plane, isotropic and steadily excited. Incoming light may be instationary; its polarization is chosen along the $z$ axis.

The vector potential now takes the form
\begin{align}
  \vek A_{{\rm TE},\vq }(\vek r, t) = \vek e_z \exp\left[\ii\qpara \vek r_{\parallel} - \ii cqt \right]\,A_{\vq}(x) \, ,
\end{align}
where $\qpara$ and $\vek r_{\parallel}$ are the in-plane components of $\vq$ and $\vek r$, and $\qperp$ and $x$ the corresponding $x$ direction components.

In contrast to, e.g., Ref.\ \onlinecite{Henneberger2009}, we will need to properly consider the coherent contribution ($\braket{\vekh A}\neq 0$) to the energy flow. In slab geometry, energy flow is only possible in the $x$ direction, and the corresponding Poynting vector component reads after insertion of the photon GFs
\begin{subequations}
  \label{eq-sx-general}
\begin{align}
  S_x(1) &= S_{x,{\rm ich}}(1) + S_{x,{\rm coh}}(1) \\
  S_{x,{\rm ich}}(1) &= \left.\frac{\hbar}{2\ii}\frac{\partial}{\partial t_1}\frac{\partial}{\partial x_2}\left( D_{zz}^>(1,2) + D_{zz}^<(1,2)\right)\right|_{2\to 1} \\
  S_{x,{\rm coh}}(1) &= \left. -\frac{1}{\mu_0}\frac{\partial}{\partial t_1}\frac{\partial}{\partial x_2}\Braket{\hat A_z(1)}\Braket{\hat A_z(2)}\right|_{2\to 1}\,.
\end{align}
\end{subequations}

The splitting of the GFs translates directly to the incoherent component of the energy flow.
For the spontaneous contribution to $S_x$, i.e., due to $D_{\rm 0,sp}\gk$ [Eq.\ \eqref{eq-d0gk-sp}], one obtains\cite{Henneberger2009}
\begin{align}
  \label{eq-sx-spontaneous}
  S_{x,\mathrm{sp}}(1) = -\frac{\hbar c^2}{2V}\sum_{\vq}\Im A_{\vq}(x) \frac{\partial}{\partial x}A_{\vq}^*(x) \,.
\end{align}

\begin{widetext}
The contribution from $D_{\rm 0,stim}$ in its general form [Eq.\ \eqref{eq-dstim-general}],
\begin{multline}
  \label{eq-sx-stim}
  S_{x,\mathrm{stim}}(1) = -\frac{\hbar c^2}{V}\sum_{\vq,\vq'}\sqrt{\frac{q}{q'}} 
 \Im \left( C_n A_{\vq}(x)\frac{\partial}{\partial x}A_{\vq'}^*(x)\exp\left [\ii(\qpara - \qpara')\vek r_{\parallel} - \ii c(q-q')t\right ]  \right. \\
 \left.+ C_a A_{\vq}(x)\frac{\partial}{\partial x}A_{\vq'}(x)\exp\left [\ii(\qpara + \qpara')\vek r_{\parallel} - \ii c(q+q')t\right ]  \right)\,,
\end{multline}
exhibits stationary terms only for $\vq' = \vq$. In the case of squeezed vacuum follows from $C_n=|\nu\ia|^2\delta\iab$ [Eq.\ \eqref{eq-d0gk-squeezed-vacuum}]
\begin{align}
  \label{eq-sx-sv}
  S_{x,\mathrm{sv}}(1) = -\frac{\hbar c^2}{V}\sum_{\vq}
\left( |\nu_{\vq}|^2 \Im A_{\vq}(x)\frac{\partial}{\partial x}A_{\vq}^*(x) \right. 
\left.+ \mu_{\vq}|\nu_{\vq}| \Im \left\{ A_{\vq}(x)\frac{\partial}{\partial x}A_{\vq}(x) \, \exp \left [2\ii(\qpara\vek r_{\parallel} - cqt) +\ii\arg \xi)\right ]  \right\} \right)\,.
\end{align}
Its homogeneous terms are stationary, the inhomogeneous terms oszillate with double frequency.

A similar analysis shows that $S_{x,{\rm coh}}$ is formally identical to $S_{x,{\rm stim}}$, so that both can be treated the same way in the following.
\begin{align}
  \label{eq-sx-coh}
  S_{x,\mathrm{coh}} = S_{x,\mathrm{stim}}[C_n \to \braket{\hat a_{\vq}}\braket{\hat a^+_{\vq'}}, C_a \to \braket{\hat a_{\vq}}\braket{\hat a_{\vq'}} ] 
\end{align}

\subsection{Scattering of incident light}
The theory so far can represent any field in the free space, since the vacuum Green function $D_0\gk$ is a sum over all wave-vectors $\vek q$, and an arbitrary state $\ket{\phi}$ is allowed. We may now address a specific situation by assuming a specific structure of the vector potential.

The following vector potential represents light incident onto both the left and the right side of the slab (\textit{symmetric setup}, with complex amplitudes $g_L$ and $g_R$), which is reflected and transmitted to the respective other side:
\begin{align}
  \label{eq-vecpot-symmetrical}
  A_{\vq}(x) = \begin{cases}
                 g_L\e^{\ii \qperp x} + (g_L r_{\vq} + g_R t_{\vq})\e^{-\ii \qperp x} & x \leq -\Lh \\
                 g_R\e^{-\ii \qperp x} + (g_R r_{\vq} + g_L t_{\vq})\e^{\ii \qperp x} & x \geq +\Lh \\
               \end{cases}\,,
\end{align}
where $r_{\vq},t_{\vq}$ are complex reflection and transmission coefficients, respectively. The vector potential $A\iq$ for coordinates inside the slab is not used in the following and does not have to be known.
This trivially covers the case of multiple incident light beams of wave vector $\vq$, which superimpose to give a new complex amplitude $g_{L/R}$. One would set $g_R = 0$ to obtain the reflection and transmission of light incident from the left alone (\textit{asymmetric setup}).

The energy flow density through the medium surface (area $F$) is
\begin{align}
  \Delta S = \frac{1}{F}\int \div \vek S \,\upd V = \frac{1}{F}\oint \vek S \, \upd f = S_x(\Lh) - S_x(-\Lh).
\end{align}
Here, $S^R = S_x(x\geq\Lh)$ describes the energy flux emitted from the right surface into the free space in parallel to the $x$ axis, while $S^L=S_x(x\leq-\Lh)$ is the left surface counterpart emitted into the negative $x$ direction.

For the calculation of $S^{L/R}$ according to Eq.\ \eqref{eq-sx-stim}, the following expressions have to be evaluated:
\begin{align}
  \left. A\iq(x)\frac{\partial}{\partial x}A\iq^*(x) \right|_{\substack{x\leq-\Lh\\x\geq+\Lh}} & = \mp\ii\qperp\left(|g_{L/R}|^2 - |g_{L/R}r\iq|^2 - |g_{R/L}t\iq|^2 + 2\ii\Im\left\{|g_{L/R}|^2(r\iq+t\iq)\exp{2\ii x}\right\}\right) \\
  \left. A\iq(x)\frac{\partial}{\partial x}A\iq(x) \right|_{\substack{x\leq-\Lh\\x\geq+\Lh}} & = \pm\ii\qperp \left(g_{L/R}^2\e^{\pm 2\ii\qperp x} - \left(g_{L/R}r\iq + g_{R/L}t\iq\right)^2\e^{\mp 2\ii\qperp x}\right)
\end{align}
The $\Im$ term in the former equation does not contribute to $S^{L/R}$ if $\Im C_n\equiv 0$, as, e.g., in (squeezed) Fock states.

Let us first analyze the spontaneous contribution (where always $|g_L|\equiv|g_R|\equiv 1$). We obtain
\begin{align}
  S^{L/R}_{\rm sp} &= \pm\frac{\hbar c^2}{V}\sum_{\vq} \qperp \frac{1}{2}\left(|g_{L/R}|^2 - |g_{L/R}r_{\vq}|^2 - |g_{R/L}t_{\vq}|^2\right) \,.
\end{align}
The balance of these energy flows is, as it should, cancelling out in the symmetric setup ($\Delta S=0$). To analyze the scattering, we set $g_R\equiv 0$ and readily reproduce the result of Ref.\ \onlinecite{Henneberger2009}: Ground state fluctuations are transmitted and reflected just as classical light, with $\left(1 - |r|^2 - |t|^2\right) = a$, where $a$ is the classical absorptivity,\cite{Henneberger2008,Richter2008b} with amplitude $\nicefrac{1}{2}$.

The $D_{0, {\rm stim}}$ contribution to $S^{L/R}$ can be obtained by straightforward calculation. We concentrate on cases with $C_{n,a} \propto \delta\iab$ and $\Im C_n\equiv 0$ for simplicity. With the help of the asymmetric setup scheme, we find that the energy flow of the incident light in general reads
\begin{align}
  S_{\rm stim}^{L,i} &= \frac{\hbar c^2}{V}\sum_{\vq} \qperp |g_L|^2 \left(C_n - |C_a|\cos\left[2\qperp x - 2 cqt + \varphi\iq\right] \right) & \varphi\iq = 2\arg g_L + \arg C_a \,,
\end{align}
while the energy flow after transmission to the right is (for reflected light accordingly)
\begin{align}
  \label{eq-sx-stim-transmitted}
  S_{\rm stim}^{R,t} &= \frac{\hbar c^2}{V}\sum_{\vq} \qperp |g_L|^2 |t_{\vq}|^2 \left(C_n - |C_a|\cos\left[2\qperp x - 2cqt + \varphi\iq\right] \right) & \varphi\iq = 2\arg g_L + \arg C_a + 2\arg t\iq \,.
\end{align}
i.e., the entire energy flow is damped by $|t\iq|^2$, and a phase delay of $2\arg t\iq$ in the instationary part is accumulated.
\end{widetext}

\subsection{Beam splitters and photon detectors}

A beam splitter, e.g., could be described following this scheme as a slab onto which light is incident from both the left and the right under an angle of 45 degrees. Then, the linearly superimposed transmitted and reflected energy flows on each side are the output channels. Any absorbing or (spatially) dispersive behavior of the beam splitter is fully accounted for by the present approach.

A photodetector measuring an electromagnetic mode converts photons in photoelectrons, hence giving rise to an electric current, called photocurrent $\hat i$. Since the Poynting vector gives the photon energy per second and square meter, it is natural to assume that the mean value of this photocurrent recorded during the (small) time interval $t,t + \Delta t$ is proportional to the normally ordered energy flow $\braket{:\hat S:}$ of the incident photons, i.e., $\braket{\hat i(t,\Delta t)} = \eta \Delta t \braket{:\hat S(t):}$, where $\eta$ is the quantum efficiency factor of the detector. Normal ordering eliminates the ground state fluctuation contribution [cf. Eq.~\eqref{eq-aa-commut}], which is usually not measured by detectors. It is equivalent to a restriction to the stimulated and coherent contributions in our approach. Hence,
\begin{align}
  \braket{:\hat S:} \equiv S_{\rm stim} + S_{\rm coh}\,,
\end{align}
is the quantity that determines the photodetection measurement outcome.

\section{Conclusion}
In this article, we show in detail how the Green function approach to the scattering of squeezed light presented in Ref.\ \onlinecite{Henneberger2009} can be generalized to light in arbitrary quantum states.

Green functions for several quantum states in the free space are derived. They can generally be separated into a ground state fluctuation and a stimulated contribution. In particular, squeezed light is addressed and the cases of general multimode squeezing, single-mode squeezing, and  two-mode squeezing are discussed. All squeezed light states are shown to result in the same squeezed vacuum contribution as a part of the stimulated contribution to the vacuum GF.

Vacuum GFs for arbitrary quantum states are constructed after normal-mode expansion of the vector potential. They are represented by normal and anormalous (inhomogeneous) terms with appropriate prefactors $C_n,C_a$ given by photon operator expectation values in the considered quantum state [Eq.\ \eqref{eq-d0gk-general}].

The normal modes propagate as classical waves in any (bounded) media system. Thus, these states are scattered no other than classical light. This conclusion is possible thanks to the exact splitting of the Green function into medium- and vacuum-induced contributions.\cite{Henneberger2009}

The same property allows to consider independently the electromagnetic energy flow contribution caused by light originating from the free space. Consequently, we develop the Poynting vector for stimulated light in its general normal-mode representation [Eqs.\ \eqref{eq-sx-general}-\eqref{eq-sx-coh}], and then discuss the description of light propagation and scattering through a slab on this basis. The advantage of this approach is the exact consideration of spatial inhomogeneities in the system and its validity for oblique incidence. Other than in the input-output formalism, there is no need for a spatial decomposition of the photon operators into incoming and outcoming, since the GFs are global functions.

After propagation through a medium, the energy flow is damped by $|t|^2$ and a phase delay of $2 \arg t$ is accumulated [Eq.\ \eqref{eq-sx-stim-transmitted}, for reflected light accordingly]. This simple result is proven here for arbitrary light states and arbitrarily absorbing or dispersive media.

Finally, it is shown how to describe some elements in an experimental setup with the present approach, for which the beam splitter serves as an example, and how the Poynting vector as a natural quantity for the energy flow relates to the outcome of measurements with photodetectors.

\begin{acknowledgments}
The authors acknowledge financial support from the {\it Deutsche Forschungsgemeinschaft} through {\it Sonder\-forsch\-ungs\-be\-reich~652}. 
\end{acknowledgments}

\end{document}